\documentclass[
superscriptaddress,
 amsmath,amssymb,
 aps, prab,
]{revtex4-2}

\usepackage{graphicx}
\graphicspath{{figures/}}
\usepackage{dcolumn}
\usepackage{bm}
\usepackage{xcolor}

\begin{document}

\preprint{APS/123-QED}

\title{\textbf{High-Harmonic Coherent Pulse Generation in a Storage Ring Using Multiple-Echo-Enabled Harmonic Generation} 
}%

\author{Weihang Liu}
\affiliation{Institute of High Energy Physics, Chinese Academy of Sciences, Beijing 100049, China}
\affiliation{Spallation Neutron Source Science Center, Dongguan 523803, China}

\author{Yu Zhao}
\affiliation{Institute of High Energy Physics, Chinese Academy of Sciences, Beijing 100049, China}
\affiliation{Spallation Neutron Source Science Center, Dongguan 523803, China}

\author{Weilun Qin}
\affiliation{Institute of High Energy Physics, Chinese Academy of Sciences, Beijing 100049, China}
\affiliation{Spallation Neutron Source Science Center, Dongguan 523803, China}

\author{Yi Jiao}
\email{jiaoyi@ihep.ac.cn}
\affiliation{Institute of High Energy Physics, Chinese Academy of Sciences, Beijing 100049, China}

\author{Xiao Li}
\email{lixiao@ihep.ac.cn}
\affiliation{Institute of High Energy Physics, Chinese Academy of Sciences, Beijing 100049, China}
\affiliation{Spallation Neutron Source Science Center, Dongguan 523803, China}

\author{Sheng Wang}
\email{wangs@ihep.ac.cn}
\affiliation{Institute of High Energy Physics, Chinese Academy of Sciences, Beijing 100049, China}
\affiliation{Spallation Neutron Source Science Center, Dongguan 523803, China}

\begin{abstract}
Fourth-generation storage-ring light sources have achieved transverse emittances approaching the diffraction limit at x-ray wavelengths, while their longitudinal coherence remains limited. {Existing laser-modulation schemes can induce strong microbunching, but most storage-ring implementations produce only one useful coherent output from a given longitudinal region during each revolution, thereby underutilizing the intrinsic multi-user capability of storage rings. We propose a multiple-echo-enabled harmonic generation (multi-EEHG) scheme that applies successive excitation--echo cycles to the same longitudinal slice of a stored bunch within one revolution, enabling coherent-radiation delivery to multiple beamlines at different wavelengths.} A general formulation of the n-stage EEHG bunching factor and a corresponding optimization procedure are derived. As an example, a triple-EEHG configuration is designed for the SAPS storage ring. Simulations demonstrate coherent radiation at multiple wavelengths with single-pulse photon numbers up to $10^9$, corresponding to an enhancement of approximately three orders of magnitude over synchrotron radiation for the same spectral bandwidth, while achieving few-meV bandwidth without a monochromator. The proposed scheme offers a scalable approach for multi-beamline coherent operation in next-generation storage-ring light sources.

\end{abstract}

\maketitle


\section{Introduction}

Storage-ring light sources (SRLSs) have become indispensable large-scale scientific facilities owing to their high brightness, stability, and operational flexibility. Over the past decade, SRLSs have advanced into the fourth generation, characterized by the adoption of multi-bend achromat (MBA) lattice designs \cite{Einfeld2014MBA}, which reduce the natural horizontal emittance to the diffraction-limited regime at x-ray wavelengths. Combined with advanced insertion devices, these developments have improved the radiation brightness by one to two orders of magnitude.

In addition to their high transverse coherence, storage rings offer unique operational advantages. The synchrotron radiation process only weakly perturbs the stored beam quality, enabling repeated reuse of the same electron bunch to serve multiple beamlines with different radiation characteristics. This feature provides high repetition rates and cost efficiency compared with linac-based light sources. However, the periodic motion of electrons in a storage ring introduces more complex beam dynamics than in linear accelerators. While MBA lattices significantly improve transverse coherence, longitudinal coherence of SRLSs remains limited.

Enhancing longitudinal coherence would enable higher spectral power density and improved temporal and energy resolution. To this end, various laser-based modulation schemes have been proposed and investigated in storage rings. These schemes generally involve two physical processes. First, an external laser interacts with the electron beam in an undulator (modulator), imposing an energy modulation. Second, the beam passes through a downstream dispersive section, where the energy modulation is converted into density modulation, leading to microbunching at high harmonics of the seed laser. The microbunched beam then emits coherent radiation in a subsequent undulator (radiator).

Among the proposed approaches, echo-enabled harmonic generation (EEHG)
\cite{Stupakov2009EchoPRL,XiangStupakov2009EEHG}, which employs two energy-modulation stages, has been extensively studied for storage-ring applications. Design studies have been reported for HEPS \cite{LiuZhouJiao2018NST}, NSLS-II \cite{YangPennYu2022SciRep}, and SOLEIL \cite{Evain2012NJP}, and EEHG-induced coherent radiation has recently been experimentally observed at DELTA \cite{KhanIPAC2024EEHGDelta}.

Angular-dispersion-induced microbunching (ADM) \cite{FengZhao2017SciRep} represents another scheme tailored to SRLSs. It exploits the ultralow vertical emittance of diffraction-limited rings to enhance the bunching factor. This concept has been investigated for several facilities, including proposals for SAPS \cite{LiuZhaoJiaoWang2023SciRep,LiuZhaoLiWangJiaoFeng2025Reverse} and other high-average-power EUV SRLSs \cite{LiFengJiang2020PRAB,Jiang2022EUVkW,LiJiangFeng2025JSR,Lu2025arXiv2511_04382}.

More generally, schemes based on transverse-longitudinal coupling \cite{Deng2021NIMA1019,LiDengPanTangChao2023PRAB} extend the ADM concept and have also attracted considerable attention. Such approaches have been discussed in the context of steady-state microbunching light sources \cite{Deng2021NatureSSMB} and related designs \cite{DengChaoHuangLiPanTang2026NST}. More recently, hybrid schemes combining ADM and transverse echo mechanisms have been proposed, opening new possibilities for coherent radiation generation in storage rings \cite{LiJiangZhang2024SciRep}.

{In parallel with these storage-ring studies, multistage EEHG concepts have been explored in linac-based free-electron lasers (FELs), primarily to push the final output toward the hard-x-ray regime. The echo-enabled staged harmonic generation scheme combines an initial EEHG stage with a downstream seeded harmonic-generation stage, using radiation from the former to interact with a fresh part of the electron bunch \cite{FengZhao2010EESHG}. Later two-stage designs used electron beams with different bunch lengths in the two stages and an intermediate monochromator to purify the first-stage radiation before it seeded the second stage \cite{ZhaoFengChenWang2016TwoBeam}. More recently, twin EEHG couples two echo stages, assigning the downstream interaction to a fresh longitudinal region to enhance the final bunching factor and generate high-power, fully coherent radiation at shorter wavelengths \cite{Ferrari2024TwinEEHG}. These linac-based schemes increase the net harmonic multiplication, but their upstream stages and intermediate or fresh electron-beam regions are used primarily to generate the final short-wavelength output.}

{Related ideas have recently been adapted or extended for storage rings. In the storage-ring cascaded-EEHG proposal of Yang \textit{et al.}, coherent radiation from the first EEHG beamline is transported as the seed of a second EEHG beamline \cite{YangYuSmaluk2023Cascaded}. In a separate configuration, an external EUV high-harmonic-generation source provides the short-wavelength seed, rather than radiation transported from an upstream beamline \cite{YangYuSmaluk2024HHGSeed}. These approaches establish important routes toward coherent emission in the tender- and hard-x-ray regimes, but their principal objective remains the final wavelength reach rather than repeated useful emission from the same spent slice.}

{A different objective was pursued by the multiplexed-emitting system proposed for an energy-recovery-linac source \cite{CaoLiuWang2023Multiplexed}. There, one pre-established microbunching structure is preserved through successive nearly isochronous transport sections and radiators, allowing one electron bunch to serve several photon beamlines. This work established the value of improving electron-bunch utilization through sequential emission. Its implementation, however, transports essentially the same density modulation between radiators and requires tight control of higher-order longitudinal transport, making it difficult to realize in a typical storage ring.}

{Taken together, these studies motivate an investigation into whether the longitudinal phase-space structure remaining after a completed echo and radiation interaction can be reused to generate a subsequent echo, thereby enabling each stage to produce a useful coherent output. We propose a multiple-echo-enabled harmonic generation (multi-EEHG) scheme that builds on this concept by cascading manipulations of the electron phase space rather than transferring the emitted radiation field between stages. At each stage, the same longitudinal slice is re-excited using its residual post-echo phase-space structure, while successive radiators generate coherent pulses at distinct harmonic wavelengths. This same-slice recycling eliminates the need for a fresh slice or interstage radiation seeding and enables a selected bunch to produce multiple useful coherent pulses in a single passage, thereby improving electron-bunch utilization while preserving the ability of a storage ring to serve multiple beamlines. The output wavelengths are coordinated rather than independently tunable, and the perturbed bunch must subsequently return toward equilibrium through radiation damping; these operational considerations are discussed below. In the following section, we formulate this mechanism for an arbitrary number of stages.}

To validate the proposed scheme, we perform numerical simulations based on the parameters of the SAPS storage ring \cite{Zhao2026SAPS}. The results demonstrate that coherent pulses at three different wavelengths can be generated using a triple-EEHG configuration, with single-pulse photon numbers up to $10^9$. After the three stages, the beam energy spread increases by approximately a factor of two; however, synchrotron radiation damping restores the equilibrium distribution within about 30 ms. Under burst-mode operation, this recovery time corresponds to an upper repetition-rate estimate of 13.5 kHz.

The remainder of this paper is organized as follows. In Sec. II, we derive the bunching factor for the multi-EEHG scheme and present the parameter-optimization conditions. Section III describes a detailed triple-EEHG design based on SAPS parameters and presents the corresponding radiation simulations. Section IV summarizes the main results and discusses the implications of the proposed scheme. The Appendix examines cumulative longitudinal wakefields, single-pass intrabeam scattering, and coherent-radiation back-reaction.

\section{ Method}
The EEHG process can be conceptually decomposed into two stages \cite{Stupakov2009EchoPRL,XiangStupakov2009EEHG}. After the first energy modulation in the upstream modulator, the beam passes through a strong dispersive section. This stage establishes a finely structured longitudinal phase-space correlation and is commonly referred to as the excitation stage.

A second energy modulation is then imposed, followed by a weaker dispersive section. This stage converts the previously induced energy modulation into density modulation and is referred to as the echo stage. Upon completion of the echo process, the microbunched beam emits coherent radiation in a downstream radiator.

{The proposed multi-EEHG scheme extends this concept by applying successive EEHG cycles to the same longitudinal slice of a stored bunch within a single revolution, as illustrated in Fig. 1. Here and throughout this work, ``the same slice'' means the electrons that have already undergone the preceding echo and radiation interaction, rather than an unused fresh longitudinal region of the bunch. In the first cycle, this slice undergoes a standard EEHG sequence and generates coherent radiation in the first radiator.}

{After the first radiation emission, the same slice enters a second strong dispersive section to initiate a new excitation stage. In contrast to the initial EEHG cycle, this stage does not require an additional upstream excitation modulation; instead, it exploits the residual post-echo longitudinal phase-space structure established by the preceding cycle. The radiation emitted by the preceding radiator is not used as the seed of the following stage.}

{A subsequent interaction with an external laser imposes an energy modulation, after which another dispersive section completes the new echo. Repetition of this excitation--echo cycle sequentially generates distinct microbunching structures within a single revolution, allowing the same electron slice to emit coherent radiation at multiple harmonic wavelengths. Thus, multi-EEHG cascades electron phase-space manipulations rather than transferring the radiation field between stages. The phase-space distribution entering each downstream stage therefore retains the cumulative effects of all preceding interactions.}

To quantitatively characterize the resulting microbunching, we now derive the corresponding analytical expression for the bunching factor.

\begin{figure}[b]
  \centering
  \includegraphics[width=0.8\columnwidth]{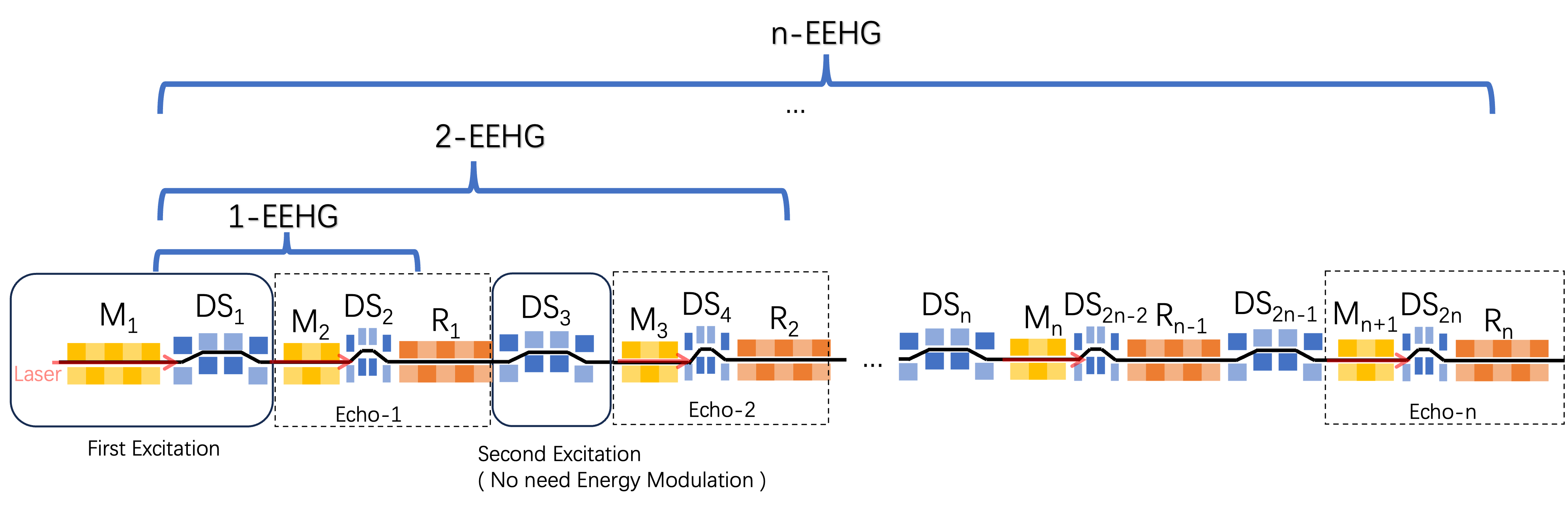}
  \caption{\label{fig:1} Schematic layout of the multi-EEHG scheme with $n$ successive stages. $\mathrm{M}_i$ ($i = 1, \ldots, n$) denote the modulators, $\mathrm{DS}_i$ the dispersive sections, and $\mathrm{R}_i$ the radiators.}
\end{figure}

In a storage ring, the natural energy spread of the electron beam is typically much larger than the free-electron laser (FEL) Pierce parameter. Moreover, the radiator length is limited to a few meters, which prevents the development of high-gain FEL amplification. As a result, radiation-induced effects remain weak. Therefore, in the theoretical analysis presented below, the influence of the radiator on the beam phase-space distribution is neglected. In the numerical simulations discussed in later sections, this effect is included for completeness.

It is also important to note that the implementation of multi-EEHG stages in a storage ring requires successive dispersive sections. In the proposed scheme, the strong dispersive sections required for each excitation stage are provided by the storage-ring arcs, which naturally supply large longitudinal dispersion. The weaker dispersive sections associated with the echo stages are realized by compact magnetic chicanes installed in the straight sections.

\subsection{Theoretical Formulation}

Let $f(p,z)$ denote the longitudinal phase-space distribution of the electron beam, where $z$ is the longitudinal coordinate and $p = \delta / \sigma_\delta$ is the normalized energy deviation. Here, $\delta$ represents the relative energy deviation, and $\sigma_\delta$ is the energy spread of the beam.

To characterize harmonic generation and microbunching, it is convenient to work in Fourier space. We define the two-dimensional Fourier transform of the longitudinal distribution as
\begin{align}
\hat{f}\left(k_z, k_p\right)=\int d z d p \times e^{i\left(z k_z+p k_p\right)} f(p,z).
\end{align}

Using this Fourier representation, the bunching factor can be directly obtained by evaluating the longitudinal component at $k_p=0$ \cite{StupakovFEL2013Coulomb}
\begin{align}
b\left(k_z\right)=\hat{f}\left(k_p=0, k_z\right).
\end{align}

We assume that the initial energy distribution of the beam is Gaussian. Since the modulation laser wavelength is much shorter than the bunch length, the longitudinal position distribution can be approximated as uniform, i.e.,
\begin{align}
    f_0(p, z)=\frac{1}{\sqrt{2 \pi} L} e^{-p^2 / 2},
\end{align}
where $L$ is the length of the electron beam.

The Fourier transform of the initial distribution is given by:
\begin{align}
    \hat{f}_0\left(k_p, k_z\right)=\sqrt{2 \pi} L \iint d p d z \exp \left[i\left(z k_z+p k_p\right)\right] f_0(p, z).
\end{align}
Substituting Eq. (3) into the above expression yields:
\begin{align}
\hat{f}_0\left(k_p, k_z\right)=e^{-k_p^2 / 2} \delta\left(k_z\right) .
\end{align}
To obtain the longitudinal distribution in Fourier space after energy modulation and to facilitate the analysis, we model the energy modulation process as:   
\begin{align}
    p=p_0+A_1 \sin \left(k_L z\right).
\end{align}
Here $p_0$ denotes the normalized energy deviation before modulation, $k_L$ is the wavenumber of the modulation laser, and $A_1$ is the amplitude of the first energy modulation. After the modulation, the distribution in Fourier space becomes:
\begin{align}
    \hat{f}_1\left(k_p, k_z\right)=\sum_{m=-\infty}^{\infty} J_{-m}\left(A_1 k_p\right) e^{-k_p^2 / 2} \delta\left(k_z-m k_L\right).
\end{align}
By comparing the right-hand side of Eq. (7) with Eq. (5), Eq. (7) can be rewritten in terms of Bessel functions and the Fourier transform of the initial distribution $\hat{f}_0$. Therefore, we obtain:
\begin{align}
    \hat{f}_1\left(k_p, k_z\right)=\sum_{m=-\infty}^{\infty} J_{-m}\left(A_1 k_p\right) \hat{f}_0\left(k_p, k_z-m k_L\right).
\end{align}
In general, the transformation of the Fourier-space distribution associated with the $n$th energy modulation can be written as:
\begin{align}
    \hat{f}_n\left(k_p, k_z\right)=\sum_{m=-\infty}^{\infty} J_{-m}\left(A_n k_p\right) \hat{f}_{n-1}\left(k_p, k_z-m k_L\right) .
\end{align}
{To derive the evolution of the Fourier-space distribution across a longitudinally dispersive section, we denote the coordinates before the section by $(z,p)$ and those after the section by $(z_1,p_1)$. The map is}
\begin{align}
    {p_1=p,\qquad z_1=z+C_1 p,}
\end{align}
{where $C_1$ denotes the strength of the first dispersive section.} The corresponding transformation of the Fourier-space distribution is
\begin{align}
    \begin{aligned}
\hat{f}_2\left(k_p, k_z\right) & =\int d z d p f_1\left(p, z-C_1 p\right) e^{i k_p p+i k_z z} \\
& =\hat{f}_1\left(k_p+k_z C_1, k_z\right).
\end{aligned}
\end{align}
In general, after a dispersive section with strength $C$, the Fourier-space density before and after the dispersive section is related by
\begin{align}
    \hat{f}_n\left(k_p, k_z\right)=\hat{f}_{n-1}\left(k_p+k_z C, k_z\right).
\end{align}
Using Eqs. (7), (9), and (12), we obtain the bunching factor for the $n$-stage EEHG process as:
\begin{align}
b(k_z) &=
\sum_{\{m_i\}=-\infty}^{+\infty}
e^{-\frac{\xi_1^2}{2}}
\prod_{i=1}^{n+1}
J_{m_i}(A_i \xi_i),
\\[6pt]
\xi_1 &= 
C_1 \!\left(k_z - \sum_{i=2}^{n+1} m_i k_L \right)
+ \sum_{j=2}^{n}
\left(C_{2j-2}+C_{2j-1}\right)
\!\left(k_z - \sum_{i=j+1}^{n+1} m_i k_L \right)
+ C_{2n} k_z,
\\[6pt]
\xi_{k+1} &= 
\xi_k
-
\left(C_{2k-2}+C_{2k-1}\right)
\left(
k_z - \sum_{i=k+1}^{n+1} m_i k_L
\right),
\\[6pt]
k_z &= \sum_{i=1}^{n+1} m_i k_L.
\end{align}
The parametric dependence of the maximum bunching on the modulation amplitudes $A_i$ and dispersive strengths $C_i$ is not immediately transparent from this expression. To clarify these dependencies, we further analyze the above result in the following.

\subsection{Numerical Analysis}

To simplify the notation, we introduce the dimensionless parameter $B_i = C_i k_L$. It can be verified that when the number of EEHG stages is 1 ($n = 1$), Eq. (13) reduces to the well-known EEHG bunching factor:
\begin{align}
    b(N)=\sum_{m_1, m_2=-\infty}^{+\infty} e^{-\frac{\xi^2}{2}} J_{m_1}\left(A_1 \xi\right) J_{m_2}\left(N A_2 B_2\right) ,
\end{align}
where $\xi$ is defined as:
\begin{align}
    \xi=B_1\left(N-m_2\right)+B_2 N.
\end{align}

As discussed in Ref. \cite{XiangStupakov2009EEHG}, for harmonic number $N$, one term in the summation of Eq. (17) dominates over all others. This so-called dominant term occurs at $m_1 = 1$ and $m_2 = N - 1$.

When $n = 2$, analytical evaluation shows that the dominant contribution to the bunching factor at harmonic $N$ can be written as:
\begin{align}
    \begin{aligned}
b_{0,1, N-1}= & J_{N-1}\left(A_3 B_4 N\right) J_1\left(A_2 \xi\right) J_0\left(A_1 \xi\right) \exp \left(-\xi^2 / 2\right).
\end{aligned}
\end{align}
where the parameter $\xi$ is given by:
\begin{align}
    \xi=B_4 N+\left(B_2+B_3\right).
\end{align}
Furthermore, for $n = 3$, the dominant term of the bunching factor at harmonic $N$ can be expressed as:
\begin{align}
    b_{0,0,1, N-1}=J_{N-1}\left(A_4 B_6 N\right) J_1\left(A_3 \xi\right) J_0\left(A_2 \xi\right) J_0\left(A_1 \xi\right) \exp \left(-\xi^2 / 2\right) .
\end{align}
where $\xi$ is given by
\begin{align}
    \xi=B_6 N+\left(B_4+B_5\right) .
\end{align}

To provide a more intuitive verification of the above analysis, we take $N = 30$ as an example and evaluate the magnitudes of the individual terms in Eq. (13) for the cases of single-, double-, and triple-stage EEHG. The results are shown in Fig. 2 and are consistent with the analytical predictions.

\begin{figure}[b]
  \centering
  \includegraphics[width=0.8\columnwidth]{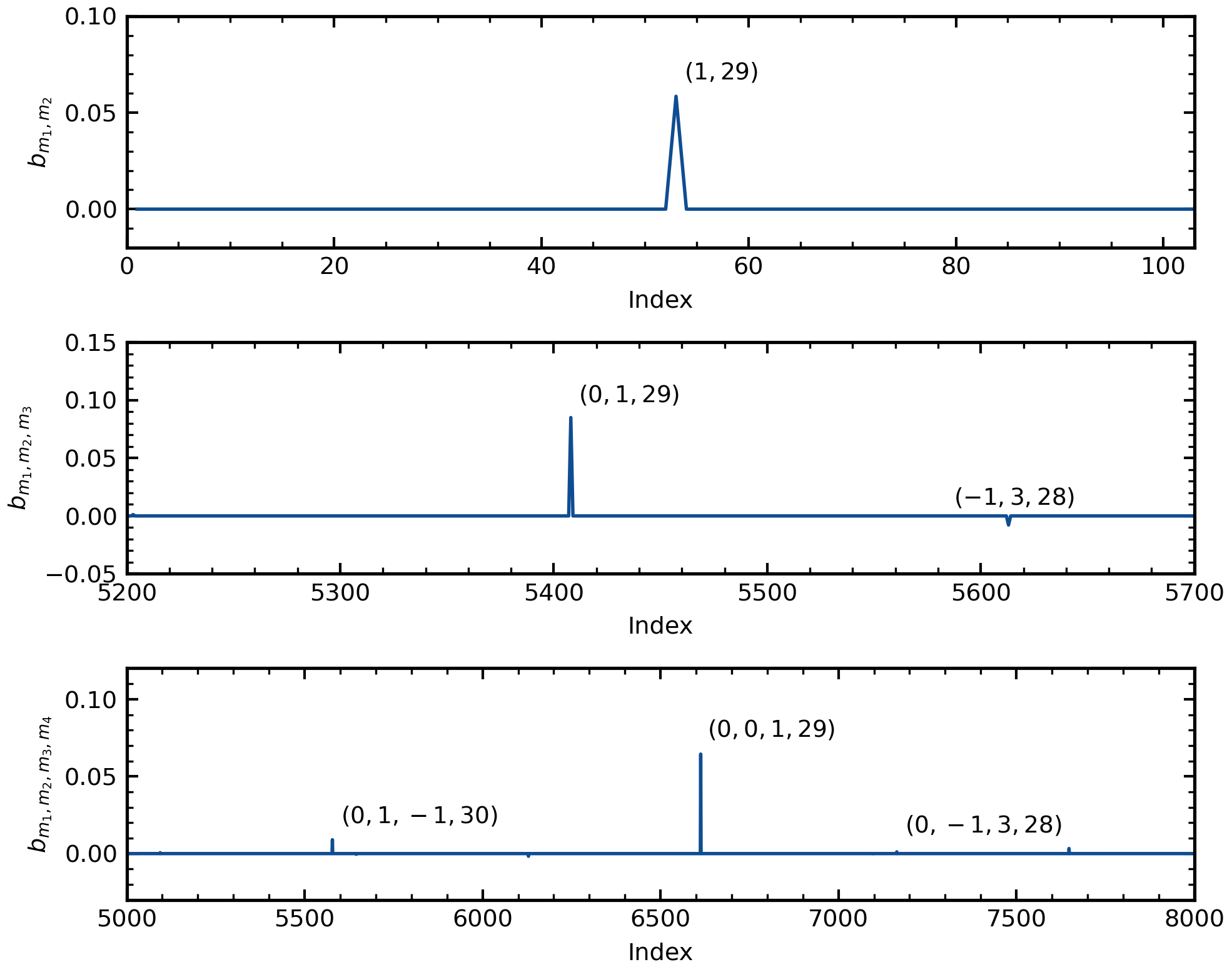}
  \caption{\label{fig:2} Contributions of individual terms to the bunching factor for $N = 30$ for 1-, 2-, and 3-stage EEHG. The labeled integer sets denote $(m_1,m_2)$, $(m_1,m_2,m_3)$, and $(m_1,m_2,m_3,m_4)$ from top to bottom.}
\end{figure}

A more detailed evaluation shows that the dominant contribution to the bunching factor at harmonic $N$ for an $n$-stage EEHG process can be written as:
\begin{align}
    b_{0,0,...,1,N-1}=J_{N-1}(A_{n+1}B_{2n}N)J_{1}\left(A_{n}\xi\right)...J_{0}\left(A_{2}\xi\right)J_{0}(A_{1}\xi)\exp(-\xi^{2}/2).
\end{align}
Here, $\xi$ is given by:
\begin{align}
    \xi=B_{2n}N+(B_{2n-1}+B_{2n-2}).
\end{align}
Using this expression, the bunching factor of the $n$-stage EEHG can be optimized systematically. The procedure is as follows. For a given set of modulation amplitudes $A_i (i = 1,2,\dots,n)$, we first determine the optimal value $\xi_{\mathrm{opt}}$ that maximizes the corresponding function of $\xi$ :
\begin{align} g(\xi)=J_1\left(A_n\xi\right)J_0\left(A_{n-1}\xi\right)...J_0\left(A_2\xi\right)J_0\left(A_1\xi\right)\exp\left(-\xi^2/2\right).
\end{align}
After obtaining $\xi_{\mathrm{opt}}$, we compute $B_{2n}$ from:
\begin{align}
    B_{2n}=\frac{\xi_{opt}-\left(B_{2n-1}+B_{2n-2}\right)}{N}.
\end{align}
Finally, by requiring the $(N-1)$ th-order Bessel function to be evaluated at its maximum, the argument $A_{n+1} B_{2n} N$ must satisfy:
\begin{align}
    A_{n+1}B_{2n}N=N-1+0.81(N-1)^{1/3}.
\end{align}
This yields the $(n+1)$th modulation amplitude $A_{n+1}$. Once $A_i$ and $B_i$ are determined, all physical parameters of the multi-EEHG configuration are fixed.

\section{Triple-EEHG Design Based on SAPS}

In this section, we apply the multi-EEHG scheme to the SAPS storage ring. SAPS is currently in the conceptual design stage, and its main parameters are listed in Table I.

We design three radiators to generate coherent radiation at wavelengths of 13.3, 8.87, and 6.65 nm, respectively. This spectral range spans part of the EUV to soft x-ray region and is of significant interest for applications such as nanoscale coherent diffraction imaging, ultrafast electron dynamics studies, and advanced lithography \cite{WuKumar2007EUVReview,Lewis2024CommunChem,HuMakZhangWeiKrausz2026LSA}. The modulation laser wavelength is chosen to be 266 nm, with the generated coherent radiation corresponding to its 20th, 30th, and 40th harmonics.

\begin{table}[!ht]
\caption{\label{tab:SAPS} Main parameters of SAPS.}
\begin{ruledtabular}
\begin{tabular}{@{}ll@{}}
Parameter & Value \\
\hline
Energy (GeV) & 3.5 \\
Energy spread (\%) & 0.1 \\
Emittance $x/y$ (pm) & 60 / 6 \\
Peak current (A) & 32 \\
Circumference (m) & 810 \\
Momentum compaction factor & $2.52 \times 10^{-5}$ \\
\end{tabular}
\end{ruledtabular}
\end{table}
Using the optimization procedure described in Sec. II, we determine the bunching factors and modulation amplitudes for the three harmonics under different parameter combinations. As indicated by Eqs. (26) and (27), larger absolute values of $B_1$, $B_3$, and $B_5$ reduce the required optimal modulation amplitudes, thereby lowering the necessary laser power. Without loss of generality, we set $B_1 = B_3 = B_5 = 2B_0$, where $B_0$ denotes the longitudinal dispersion of one arc, and choose $A_1 = 1.5$.

For the three harmonics, six possible configurations arise, which we refer to as different pattern. As shown in Fig. 3, the corresponding bunching factors vary among these patterns. To achieve comparable and sufficiently strong radiation at all three wavelengths, we evaluate the average bunching factor of the three harmonics. The results indicate that the descending order configuration $(40, 30, 20)$ provides the most favorable performance.

\begin{figure}[!ht]
  \centering
  \includegraphics[width=0.8\columnwidth]{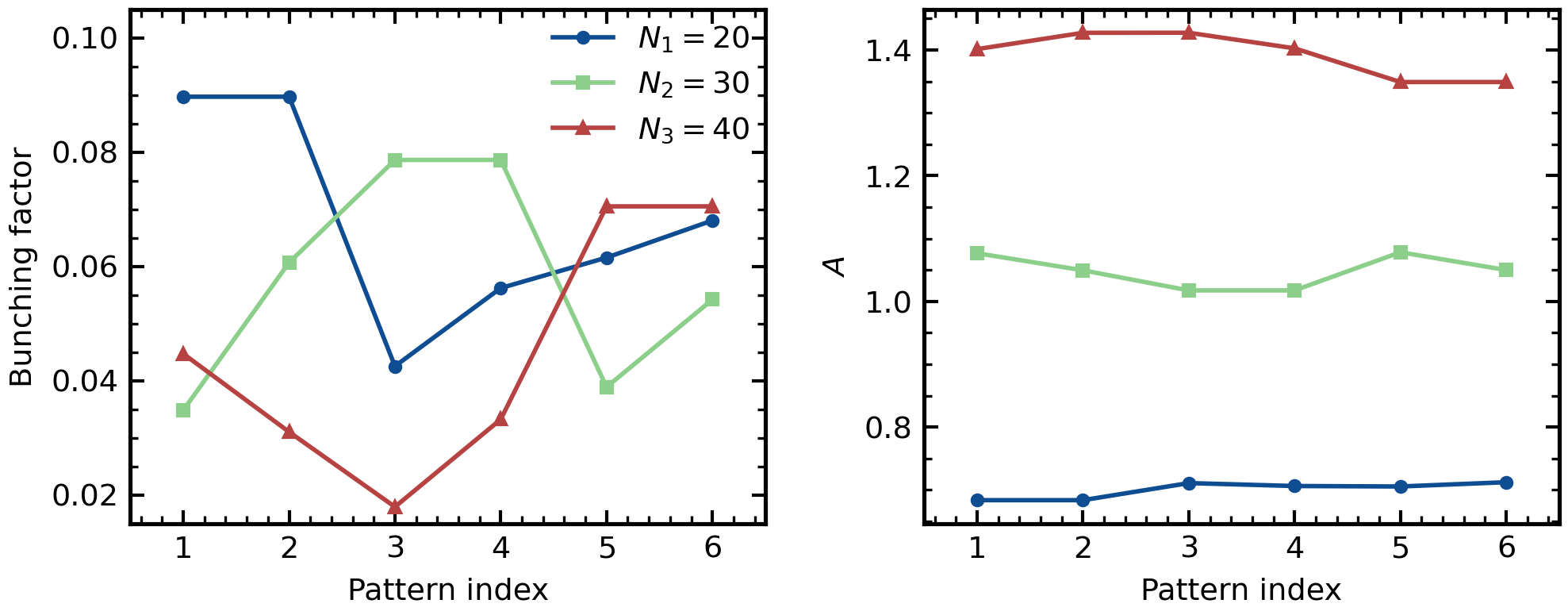}
  \caption{\label{fig:3} Bunching factors and corresponding modulation amplitudes for the triple-EEHG scheme under different harmonic configurations $(N_1,N_2,N_3)$. Pattern indices 1–6 correspond to $(N_1,N_2,N_3)=(20,30,40)$, $(20,40,30)$, $(30,20,40)$, $(30,40,20)$, $(40,20,30)$, and $(40,30,20)$.
  }
\end{figure}

\begin{table}[!ht]
\caption{\label{tab:tripleEEHG}
Parameters of the triple-EEHG configuration based on SAPS.}
\begin{ruledtabular}
\begin{tabular}{@{}l r r r r r r@{}}
 & 1 & 2 & 3 & 4 & 5 & 6 \\
\hline
$A_i$ & 1.50 & 1.35 & 1.05 & 0.70 & -- & -- \\
$B_i$ & $-30.16$ & 0.77 & $-30.16$ & 1.00 & $-30.16$ & 1.49 \\
$R_{56,i}$ ($\mu$m) 
& $-1277$ & 32.8 & $-1277$ & 42.3 & $-1277$ & 62.9 \\
Laser power (GW) 
& 12.9 & 10.6 & 6.3 & 2.9 & -- & -- \\
\end{tabular}
\end{ruledtabular}
\end{table}

The optimal parameters obtained from the theoretical analysis are summarized in Table II. The predicted bunching factors at the 20th, 30th, and 40th harmonics are 6.8\%, 5.4\%, and 7.1\%, respectively. It can be observed that the required modulation amplitudes decrease progressively from the first to the fourth stage.

{To improve laser-power utilization, a single seed pulse could, in principle, be routed sequentially from the first modulator to the second, third, and fourth modulators through an optical transport line, with attenuators controlling the power delivered at each interaction. Dielectric multilayer mirrors can provide reflectivities exceeding 90\% at wavelengths above 200~nm \cite{edmund_mirror}, thereby limiting transport losses. High reflectivity alone, however, does not guarantee the feasibility of pulse reuse. Between each pair of successive modulators, the minimum realizable optical path length must not exceed the corresponding electron reference-orbit length; moreover, sufficient adjustable delay must remain after accommodating accelerator components, radiation shielding, and photon extraction. The viability of this architecture must therefore be established through a lattice-specific geometric and timing analysis. If these constraints cannot be satisfied, independently synchronized seed lasers provide an alternative. The simulations presented below focus exclusively on the beam-dynamics performance of the multi-EEHG mechanism and assume prescribed laser powers and ideal laser--electron synchronization; they do not validate the engineering feasibility of either laser architecture.}

\begin{figure}[!ht]
  \centering
  \includegraphics[width=\columnwidth]{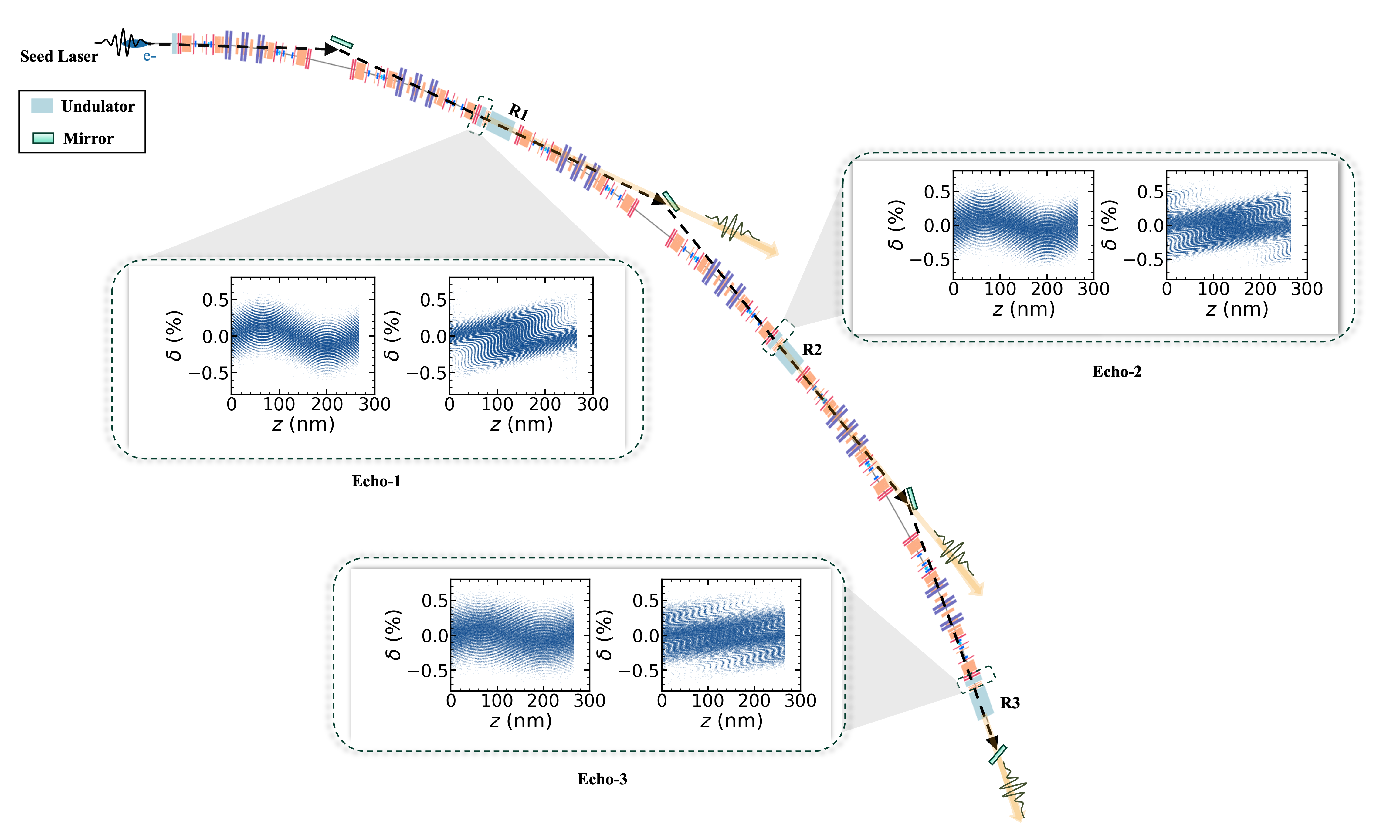}
  \caption{\label{fig:4} Schematic layout of the triple-EEHG configuration and the corresponding longitudinal phase-space evolution for the successive echo processes.}
\end{figure}

All four modulators employ undulators with a period length of 13.3 cm and a total length of approximately 2 m. To generate radiation at different wavelengths, three radiators with distinct undulator periods of 4 cm, 4.5 cm, and 5 cm are used, each with a total length of 3.6 m. The modulation laser has a pulse duration of approximately 424 fs.

The overall layout of the triple-EEHG scheme in SAPS is illustrated in Fig. 4. The electron bunch first undergoes energy modulation in a modulator installed in a straight section, and then passes through two adjacent arcs that provide the strong dispersion required for the first excitation stage. The first echo process is completed in a downstream straight section, where coherent radiation is generated in the corresponding radiator.

Subsequently, the same bunch proceeds through another pair of adjacent arcs and a straight section to realize the second EEHG stage, followed by a further set of arcs and a straight section to complete the third EEHG stage.

Beam dynamics are simulated using the code ELEGANT \cite{Borland2000ELEGANT}, including coherent synchrotron radiation, incoherent synchrotron radiation, and nonlinear effects. The radiation–beam interaction in the modulators and radiators is simulated using GENESIS \cite{Reiche1999GENESIS} and incorporated into the tracking. As shown in Fig. 4, after each echo stage, the longitudinal phase-space distribution exhibits the characteristic fine structures of EEHG. {Further analysis of the relevant longitudinal effects is provided in Appendix~\ref{app:collective}.}

It is worth noting that the strengths of the chicane dispersive sections, $B_2$, $B_4$, and $B_6$, can be further optimized by directly evaluating the radiation power to achieve enhanced output performance. Using the time-independent mode of GENESIS, we performed parameter scans of the radiation power as a function of the corresponding $B_i$.

The results are presented in Fig. 5. From these scans, the optimal chicane strengths correspond to $R_{56}$ values of 30.78 $\mu$m, 38.53 $\mu$m, and 58.20 $\mu$m, respectively.

\begin{figure}[!ht]
  \centering
  \includegraphics[width=0.8\columnwidth]{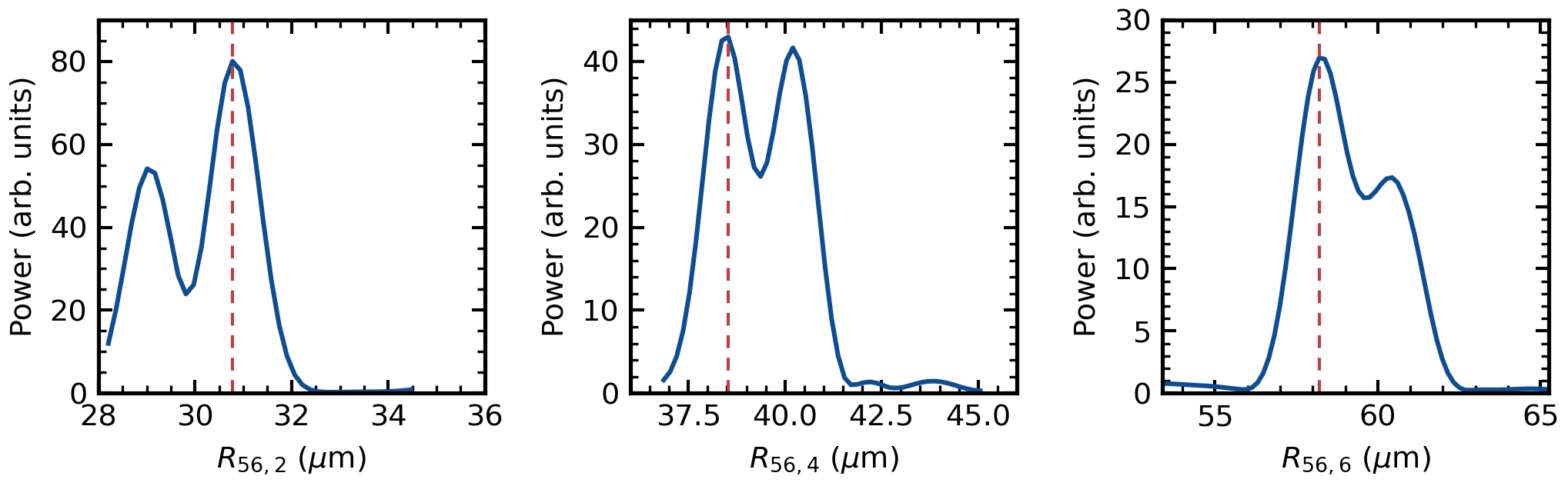}
  \caption{\label{fig:5} Dependence of the radiation power on the longitudinal dispersion strength for the first, second, and third echo stages (from left to right).}
\end{figure}

After determining the optimal chicane strengths, we perform time-dependent simulations using GENESIS to obtain the radiation characteristics for the three harmonics.

\begin{figure}[!ht]
  \centering
  \includegraphics[width=0.8\columnwidth]{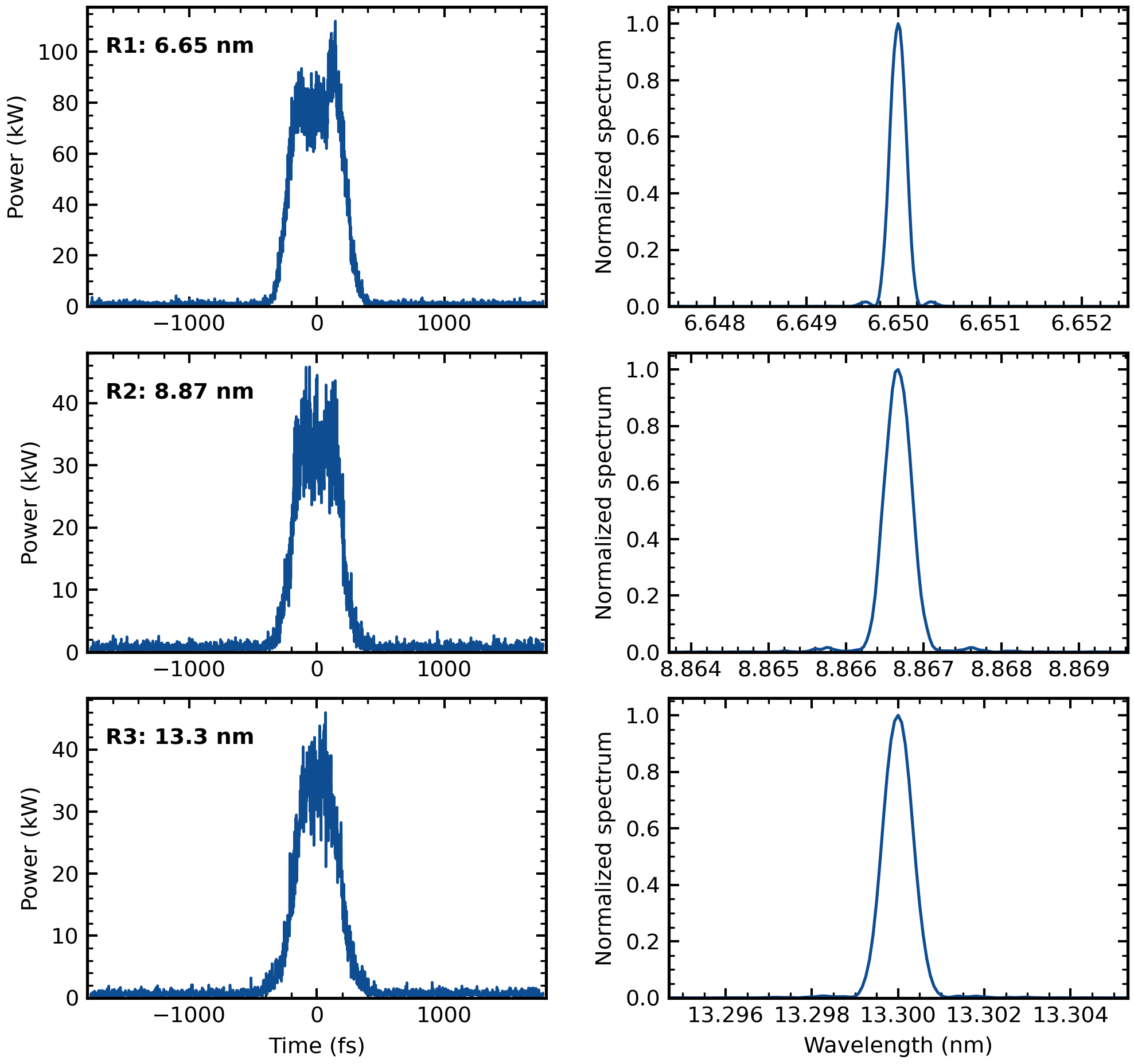}
  \caption{\label{fig:6} Radiation power profiles and corresponding spectra for the radiators associated with the first, second, and third echo stages (from top to bottom).}
\end{figure}

As shown in Fig. 6, the peak radiation power at each wavelength exceeds 10 kW, with a pulse duration of approximately 400 fs. The corresponding single-pulse photon number can reache $10^9$. The spectral bandwidth for the different wavelengths is about 6 meV. For the same spectral bandwidth, the photon number per pulse is approximately three orders of magnitude higher than that of conventional synchrotron radiation from a single bunch.

{
After three successive echo stages, the rms energy spread increases from its equilibrium value $\sigma_{\delta,0}=0.1\%$ to approximately $\sigma_{\delta,\mathrm m}=0.2\%$.  Recovery requires about three longitudinal damping times, corresponding to roughly $T_{\mathrm{rec}}=30$~ms, after which the same bunch can be remodulated.  To quantify the associated influence on conventional synchrotron-radiation beamlines, we describe the recovery of the energy-spread variance by
\begin{equation}
 \sigma_\delta^2(t)=\sigma_{\delta,0}^2+
 \left(\sigma_{\delta,\mathrm m}^2-\sigma_{\delta,0}^2\right)
 \exp\left(-\frac{2t}{\tau_z}\right),
 \label{eq:energy_spread_recovery}
\end{equation}
with $\tau_z=10$~ms.  A similar brightness-based assessment of the effect of a laser-perturbed bunch on other synchrotron-radiation beamlines was performed in Ref.~\cite{LiuZhaoJiaoWang2023SciRep}.  Following the same method, we evaluate the spectral brightness using the universal energy-spread correction of Tanaka and Kitamura \cite{TanakaKitamura2009Brilliance}.  We use the parameters of a representative SAPS in-vacuum undulator (IVU), $\lambda_u=18.5$~mm, $N_u=258$, and $B_{\max}=1.1$~T \cite{YangJiaoLiu2026IDCompensation}, and consider harmonics up to the fifth.  
}

\begin{figure}[!ht]
  \centering
  \includegraphics[width=\columnwidth]{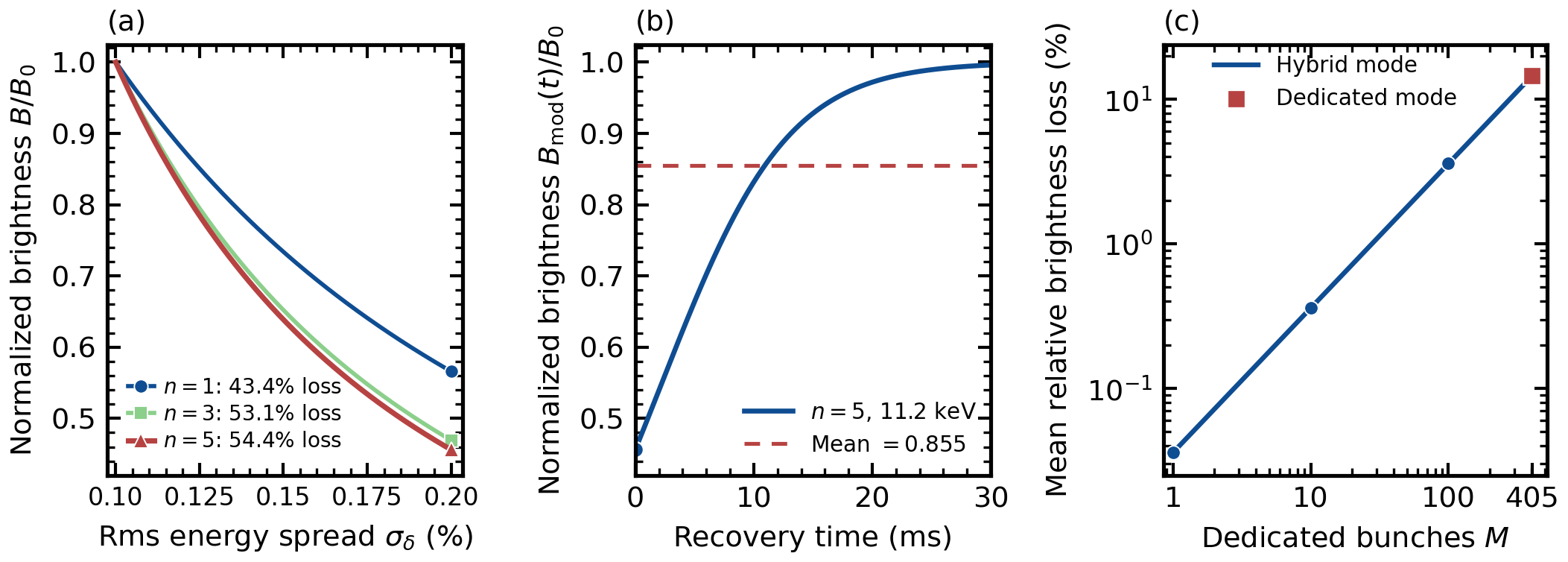}
  \caption{\label{fig:synchrotron_brightness_impact}{Influence of the post-multi-EEHG energy spread on the normalized spectral brightness of a representative IVU18.5 beamline.  (a) Energy-spread sensitivity of the first, third, and fifth harmonics at the maximum-field working point.  (b) Recovery of the fifth-harmonic brightness at 11.2~keV during one 30-ms damping interval; the dashed line denotes its time average.  (c) Mean brightness loss seen by the representative IVU18.5 beamline as a function of the number $M$ of designated multi-EEHG bunches in a 405-bunch filling pattern.}}
\end{figure}

{
Figure~\ref{fig:synchrotron_brightness_impact}(a) shows that, among the three IVU harmonics considered, the fifth harmonic is the most sensitive to energy spread: increasing $\sigma_\delta$ from $0.1\%$ to $0.2\%$ reduces its instantaneous brightness by $54.4\%$.  This penalty decreases as the perturbed bunch recovers through radiation damping.  As shown in Fig.~\ref{fig:synchrotron_brightness_impact}(b), the fifth-harmonic brightness averaged over the 30-ms recovery interval is $\langle B_{\mathrm{mod}}/B_0\rangle_t=0.855$, where $B_0$ denotes the equilibrium brightness.

During each multi-EEHG passage, a selected bunch sequentially serves the three coherent-radiation beamlines.  Two operating modes may therefore be considered.  In the dedicated coherent-radiation mode, all 405 bunches in the assumed 90\% filling pattern undergo multi-EEHG in turn, providing a maximum coherent-pulse repetition rate of approximately 13.5~kHz.  Because every bunch is repeatedly subjected to multi-EEHG and subsequent radiation damping, a synchrotron-radiation beamline would experience the 14.54\% cycle-averaged brightness loss shown in Fig.~\ref{fig:synchrotron_brightness_impact}(b).  Such a loss may be unacceptable to most synchrotron-radiation users, making this mode more suitable for dedicated coherent-radiation operation.

Alternatively, in a hybrid or special-bunch mode, only $M$ designated bunches, with $1\leq M<405$, undergo multi-EEHG, resulting in a coherent-pulse repetition rate of approximately $M/T_{\mathrm{rec}}$.  A synchrotron-radiation beamline receives radiation from both equilibrium and perturbed bunches.  Assuming equal bunch charges, its mean normalized brightness is
\begin{equation}
 \frac{\overline{B}(M)}{B_0}
 =1-\frac{M}{405}\left(1-
 \left\langle\frac{B_{\mathrm{mod}}(t)}{B_0}\right\rangle_t\right).
 \label{eq:hybrid_brightness_average}
\end{equation}
As shown in Fig.~\ref{fig:synchrotron_brightness_impact}(c), the mean brightness losses are $0.0359\%$, $0.359\%$, and $3.59\%$ for $M=1$, 10, and 100, respectively.  Thus, the hybrid mode can substantially reduce the impact of multi-EEHG operation on synchrotron-radiation brightness.
}

\begin{figure}[!ht]
\centering

\includegraphics[width=0.8\columnwidth]{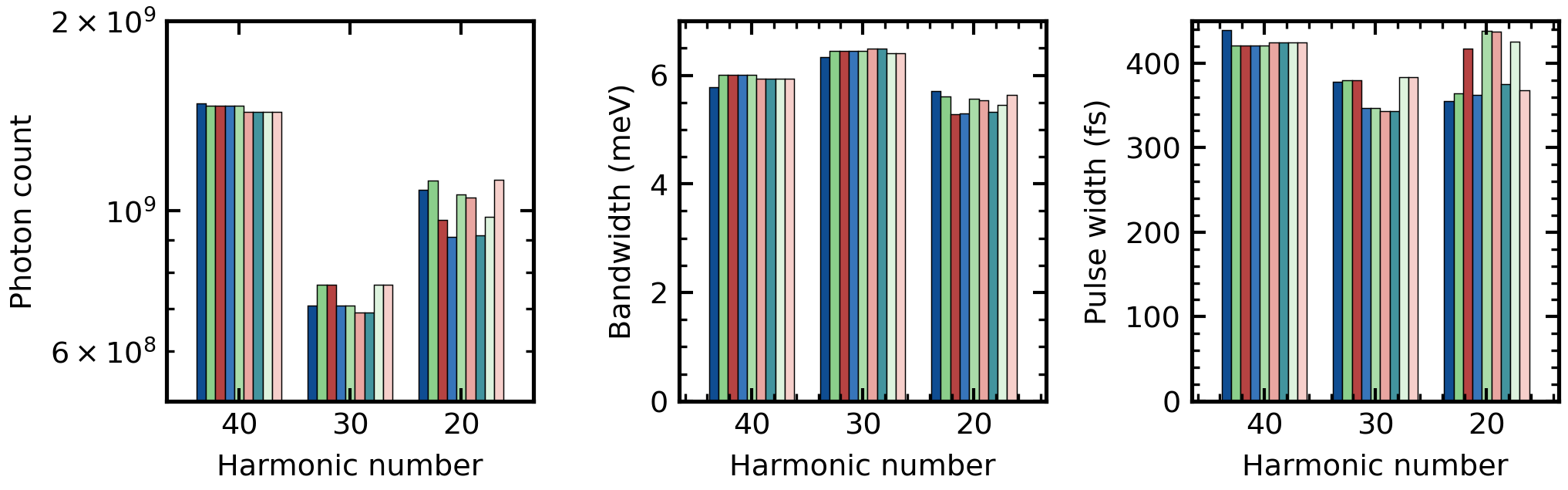}

\vspace{6pt}

\includegraphics[width=0.8\columnwidth]{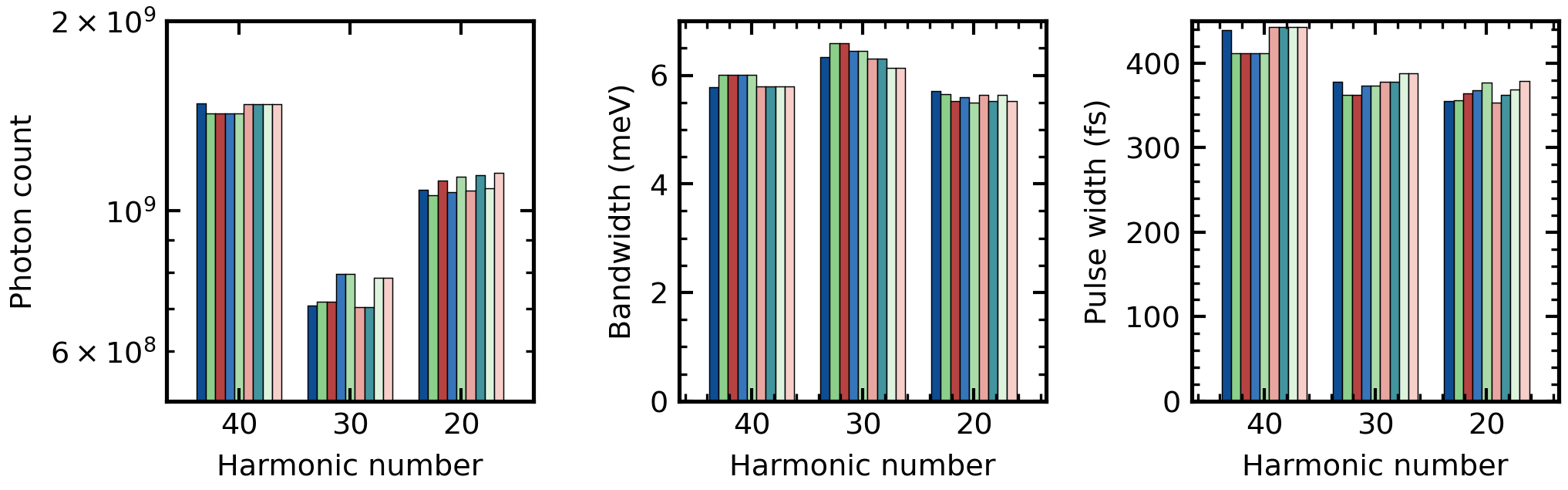}

\caption{\label{fig:fig7}
Radiation performance of different harmonics under laser parameter deviations. The top row corresponds to timing offsets of the modulation laser, and the bottom row to relative intensity variations.
}
\end{figure}

{The cumulative dependence of each downstream stage on the phase-space distribution established by all preceding interactions introduces correlations among the output characteristics, limiting the extent to which they can be adjusted independently. Fluctuations in the modulation-laser intensity and timing relative to the electron bunch may therefore affect the radiation performance across the multistage sequence. To quantify these effects, we assume a maximum timing offset of 100 fs and a maximum relative intensity variation of 1\% at each echo stage.}

Since three echo stages are involved and each stage is subject to two independent parameter deviations (timing and intensity), nine cases are considered for each stage. {Figure~\ref{fig:fig7}} shows the resulting radiation performance variations, with the top row corresponding to timing offsets only and the bottom row to intensity fluctuations only. The results indicate that both timing and intensity deviations lead to relatively small variations in radiation performance. The spectral bandwidth and pulse duration are the least sensitive to these deviations, with variations below 5\% for all harmonics. Although the single-pulse photon number shows a stronger sensitivity, particularly for lower harmonic numbers, the variation remains limited to about 10\%.

\section{Conclusion}
{We have proposed a multiple-echo-enabled harmonic generation (multi-EEHG) scheme that generates multiple coherent pulses from a selected electron-bunch passage in a storage ring by successively reusing the longitudinal phase-space structure of the same electron slice.  A general $n$-stage bunching formalism and a corresponding optimization procedure have been developed, providing a scalable framework for coordinated multiwavelength operation.  Multi-EEHG thereby improves the utilization of the modulated bunch and extends high-harmonic coherent-radiation generation to multiple beamlines.}

{Numerical simulations using the SAPS storage-ring parameters demonstrate coherent emission at multiple wavelengths.  For the same spectral bandwidth, the single-pulse photon numbers exceed those of conventional synchrotron radiation by approximately three orders of magnitude, while few-meV bandwidths are achieved without monochromation.  The analyses of collective effects and coherent-radiation back-reaction further show that, under the assumed SAPS conditions, cumulative longitudinal wakefields, single-pass intrabeam scattering, and radiator back-reaction remain sufficiently weak to preserve the phase-space structure required for subsequent echoes.  These results demonstrate the potential of multi-EEHG to provide high-brightness, narrow-band coherent radiation to multiple beamlines in storage-ring light sources.}

{The output wavelengths and pulse characteristics of different stages are correlated and therefore cannot, in general, be tuned fully independently.  Nevertheless, the laser-parameter sensitivity study shows that, for the timing and intensity deviations considered here, variations in spectral bandwidth and pulse duration remain below 5\%, while changes in the single-pulse photon number are limited to about 10\%.  Future studies should pursue application-oriented joint optimization of these outputs to meet the specific requirements of multiple beamlines.  Experimental implementation will also require facility-specific tolerance studies incorporating realistic lattice and hardware imperfections, impedance distributions, and laser fluctuations.}

{As a natural extension of EEHG, multi-EEHG retains substantial scalability and flexibility.  Concepts developed for conventional EEHG, including few-cycle laser modulation for attosecond pulse generation \cite{ZholentsPenn2010NIMA,Hwang2020SciRep} and vortex-laser modulation for producing radiation carrying orbital angular momentum \cite{HemsingMarinelli2012PRL}, could in principle be incorporated into the multi-EEHG framework.  These possibilities establish multi-EEHG as a promising route toward versatile and application-tailored coherent light sources in next-generation storage-ring facilities.}

\appendix
{
\section{Collective effects and coherent-radiation back-reaction in multi-EEHG}
\label{app:collective}

Preservation of the fine longitudinal phase-space structure inherited from each preceding echo--radiator stage is essential for downstream echo formation in multi-EEHG.  Accordingly, we examine three effects that accumulate over a single passage: longitudinal wakefields, intrabeam scattering (IBS), and coherent-radiation back-reaction.

\subsection{Cumulative longitudinal-wakefield effects}

For a line-charge distribution $\lambda(z)$ normalized to the bunch charge $Q$, a longitudinal wake source at position $s$ produces
\begin{equation}
 V_{\mathrm w}(z;s)=-\int_{-\infty}^{z}
 W_{\parallel}(z-z';s)\lambda(z')\,\mathrm dz',
 \qquad
 \Delta\delta_{\mathrm w}(z;s)=\frac{eV_{\mathrm w}(z;s)}{E_0},
 \label{eq:wake_convolution}
\end{equation}
where $W_{\parallel}$ is the causal longitudinal wake function and $E_0$ is the beam energy.  The distributed-impedance contribution estimated below is treated as varying on the bunch-envelope scale.  The approximately 400~fs radiation-producing core spans about 450 periods of the 266~nm modulation laser, so the wake-induced energy variation is locally slow on the optical scale.  This is the same scale separation used by Hemsing, under which an additional energy structure enters primarily through the longitudinal phase of the complex EEHG bunching factor \cite{Hemsing2018BunchingPhase}.

An energy perturbation affects the final bunching phase through the dispersive transport downstream of its point of origin.  If it is generated upstream of a complete echo stage, the contributions from the strong and weak dispersive sections approximately cancel for the optimized EEHG dispersions.  If it is generated within the strong dispersive section, however, only the remaining fraction of that section contributes together with the downstream weak dispersive section; the cancellation is therefore incomplete, and the phase sensitivity becomes location dependent.  Because the multi-EEHG map in Sec.~II is constructed from the same Fourier-space transformations for energy modulation and longitudinal shear, this phase-propagation method can be applied stage by stage.  The resulting location-dependent coefficient is derived below.

To make the position-dependent coefficient explicit, consider a specified radiator $R_j$, whose target harmonic number is denoted by $N_j$.  Setting $n=j$ in Eq.~(13), we select a Fourier branch $\boldsymbol m=(m_1,\ldots,m_{j+1})$ satisfying $\sum_{a=1}^{j+1}m_a=N_j$.  These $m_a$ are the same Bessel orders used in Sec.~II.  Let $k_{z,\ell}^{(j)}$ denote the longitudinal Fourier coordinate of this branch in dispersive section $D_\ell$.  Equation~(13) gives the directly evaluable relation
\begin{equation}
 k_{z,\ell}^{(j)}=k_L
 \sum_{a=1}^{\lfloor\ell/2\rfloor+1}m_a,
 \qquad \ell=1,\ldots,2j.
 \label{eq:wake_kz_branch}
\end{equation}
Equivalently, starting from $k_{z,2j}^{(j)}=N_jk_L$ at $R_j$, crossing modulator $M_{r+1}$ backward gives $k_{z,2r-1}^{(j)}=k_{z,2r}^{(j)}-m_{r+1}k_L$, while $k_{z,2r-2}^{(j)}=k_{z,2r-1}^{(j)}$ when no modulation lies between the two sections.  Define $C_\ell^{\mathrm{down}}(s)$ as the portion of $C_\ell$ downstream of the kick: it equals $C_\ell$ if $s$ is upstream of $D_\ell$, the remaining dispersion from $s$ to the exit if $s$ lies inside $D_\ell$, and zero if $s$ is downstream.  The dimensionless phase-sensitivity coefficient is then
\begin{equation}
 \chi_j(s)=\sum_{\ell=1}^{2j}
 C_\ell^{\mathrm{down}}(s)k_{z,\ell}^{(j)}
 =\sum_{\ell=1}^{2j}B_\ell^{\mathrm{down}}(s)
 \sum_{a=1}^{\lfloor\ell/2\rfloor+1}m_a.
 \label{eq:wake_chi_definition}
\end{equation}

With $\Delta p_{\mathrm w}=\Delta\delta_{\mathrm w}/\sigma_\delta$, a set of localized wake kicks at $s_a$ therefore produces, using the Fourier convention of Sec.~II,
\begin{equation}
 \psi_{\mathrm w,j}(z)=\sum_a\chi_j(s_a)
 \Delta p_{\mathrm w}(z;s_a),\qquad
 b_j(k)\simeq\int dz\,b_{j,0}(z;k)
 e^{i\psi_{\mathrm w,j}(z)}.
 \label{eq:wake_phase}
\end{equation}
For a distributed wake source, the discrete sum is replaced by a longitudinal integral over the differential kicks along the beamline.  For a single EEHG stage, Eq.~(\ref{eq:wake_phase}) reduces to the result of Hemsing and explicitly accounts for the near cancellation between the phase contributions of the optimized strong and weak dispersive sections.

For the present dominant branches,
\begin{equation}
 \begin{aligned}
 R_1:\quad \boldsymbol m&=(1,N_1-1),
 &\frac{\boldsymbol k_z^{(1)}}{k_L}&=(1,N_1),\\
 R_2:\quad \boldsymbol m&=(0,1,N_2-1),
 &\frac{\boldsymbol k_z^{(2)}}{k_L}&=(0,1,1,N_2),\\
 R_3:\quad \boldsymbol m&=(0,0,1,N_3-1),
 &\frac{\boldsymbol k_z^{(3)}}{k_L}&=(0,0,0,1,1,N_3).
 \end{aligned}
 \label{eq:wake_dominant_branches}
\end{equation}
Consequently, a kick produced at position $s$ inside the current strong section $D_{2j-1}$ has
\begin{equation}
 \chi_j(s)=B_{2j-1}^{\mathrm{rem}}(s)+N_jB_{2j},
 \qquad s\in D_{2j-1},
 \label{eq:wake_current_arc_weight}
\end{equation}
where $B_{2j-1}^{\mathrm{rem}}(s)$ is the remaining dispersion from $s$ to the exit of the strong section.  For wake kicks located upstream of a complete echo transport, the optimized strong- and weak-section dispersions give nearly canceling phase contributions, leaving only a small residual weight.  For $R_1$, the coefficient at the entrance of the first strong section is $\xi_1$; for the downstream radiators $R_2$ and $R_3$, contributions originating in preceding arc pairs enter with the coefficients $\xi_2$ and $\xi_3$, respectively.  For the optimized parameters, these coefficients are
\begin{equation}
 \xi_1=B_1+N_1B_2=0.64,\quad
 \xi_2=B_2+B_3+N_2B_4=0.61,\quad
 \xi_3=B_4+B_5+N_3B_6=0.64.
 \label{eq:wake_echo_scaling}
\end{equation}
Across the current strong section, Eq.~(\ref{eq:wake_current_arc_weight}) varies from the near-cancelled entrance value to $N_jB_{2j}$ at its exit.  Hence $|\chi_j|$ is conservatively bounded by $|N_jB_{2j}|=(30.8,30.0,29.8)$, whereas earlier-stage contributions retain the much smaller complete-echo weights.  Wake-induced energy loss therefore accumulates from one arc pair to the next, but its bunching-phase contribution accumulates with the location-dependent coefficient in Eq.~(\ref{eq:wake_chi_definition}).

For a numerical scale, the reported full-ring loss factor $\kappa_l=32.2$~V/pC \cite{Li2026SAPSCollective} gives a mean energy loss of $\kappa_lQ_{\mathrm{sim}}=3.66$~keV per electron for $Q_{\mathrm{sim}}=113.57$~pC.  Assuming that this impedance is distributed uniformly over the 32 arc units, each pair of arcs contributes $\Delta E_{\mathrm{pair}}\simeq(2/32)\times3.66=0.229$~keV.  In the normalization of Eq.~(\ref{eq:wake_phase}), the magnitude of the wake kick assigned to each pair is $|\Delta p_{\mathrm w}|=\Delta E_{\mathrm{pair}}/\sigma_E=6.54\times10^{-5}$ for $\sigma_E=3.5$~MeV.  If the full mean loss of the current pair is treated as an uncorrectable variation with its maximum local weight, and all preceding pairs are added coherently with the complete-echo weight, then
\begin{equation}
 |\psi_{\mathrm w,j}|\le|\Delta p_{\mathrm w}|
 \left[|N_jB_{2j}|+(j-1)|\xi_j|\right],
 \qquad j=1,2,3.
 \label{eq:wake_phase_bound_rule}
\end{equation}
This conservative upper-bound estimate gives
\begin{equation}
 |\psi_{\mathrm w,1}|<2.01\times10^{-3},\quad
 |\psi_{\mathrm w,2}|<2.00\times10^{-3},\quad
 |\psi_{\mathrm w,3}|<2.03\times10^{-3}\ \mathrm{rad}.
 \label{eq:wake_phase_bounds}
\end{equation}
Even if the largest value in Eq.~(\ref{eq:wake_phase_bounds}) is treated as a Gaussian rms phase variation, the bunching-magnitude reduction is only $1-\exp[-(2.03\times10^{-3})^2/2]=2.06\times10^{-6}$.  The resulting phase shift and bunching reduction are therefore too small to affect the multi-EEHG process appreciably within the stated uniform-impedance model.

\subsection{Single-pass intrabeam-scattering effects}

IBS produces stochastic diffusion in six-dimensional phase space and may broaden the fine energy bands required by a subsequent echo.  Because the multi-EEHG process is completed within one revolution, we evaluate the diffusion accumulated through the two storage-ring arcs in each strong excitation section.  This single-pass contribution was calculated with the Bjorken--Mtingwa model implemented by the \textsc{ELEGANT} \texttt{IBSCATTER} element \cite{Borland2000ELEGANT,StupakovFEL2013Coulomb}.  Two simulations of the complete three-stage sequence were initialized with the same macroparticle distribution.  In the IBS case, \texttt{IBSCATTER} was enabled in each of the six physical arcs; it was disabled in the reference case.  At each stage, the six-dimensional distribution obtained from \textsc{ELEGANT} tracking was used as the input to the corresponding GENESIS radiator calculation, and the radiator-exit particles were subsequently passed to the next \textsc{ELEGANT} transport section.  All other lattice, ISR, CSR, seed-laser, and GENESIS settings were identical, so the difference between the two simulations isolates the incremental effect of single-pass IBS.

Figure~\ref{fig:ibs_effects} compares the cases with and without IBS.  The upper panels show the energy bands after transport through the first pair of physical arcs and before the first short chicane, while the lower panels show the radiation from the complete downstream chains.  At this observation point, the accumulated IBS-induced energy displacement was evaluated particle by particle as
\begin{equation}
 \sigma_{\Delta\delta,\mathrm{IBS}}
 =\left\langle
 \left(\delta_i^{\mathrm{IBS}}-\delta_i^{\mathrm{no\text{-}IBS}}\right)^2
 \right\rangle_i^{1/2}
 =6.46\times10^{-6},
 \label{eq:ibs_energy_displacement}
\end{equation}
corresponding to $22.6$~keV at 3.5~GeV.  This rms displacement is only $7.0\%$ of the minimum center-to-center separation between adjacent energy bands, $9.26\times10^{-5}$ ($324$~keV), and is therefore too small to cause appreciable mixing of neighboring bands, consistent with the preserved band structure in Fig.~\ref{fig:ibs_effects}.

Consistent with the preserved energy-band structure, IBS produces only few-percent changes in the downstream microbunching.  Relative to the no-IBS case, the peak slice bunching factors at the entrances to $R_1$--$R_3$ are reduced by $1.18\%$, $1.06\%$, and $0.76\%$, respectively.

\begin{figure*}[!t]
 \centering
 \includegraphics[width=0.86\textwidth]{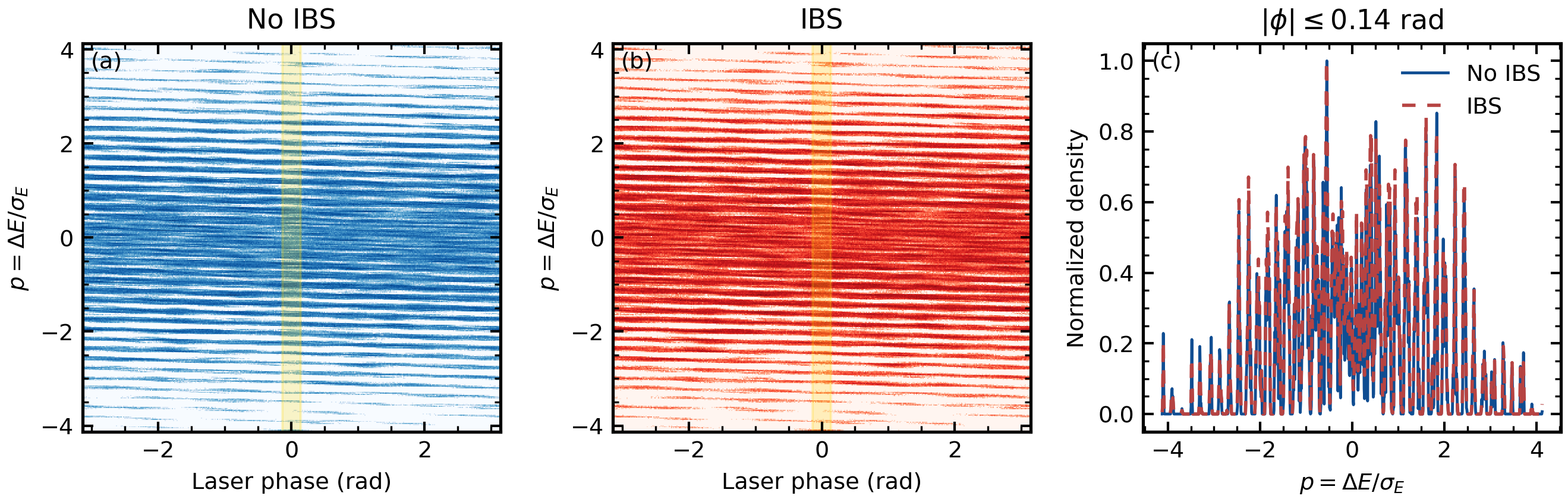}\\[3pt]
 \includegraphics[width=0.68\textwidth]{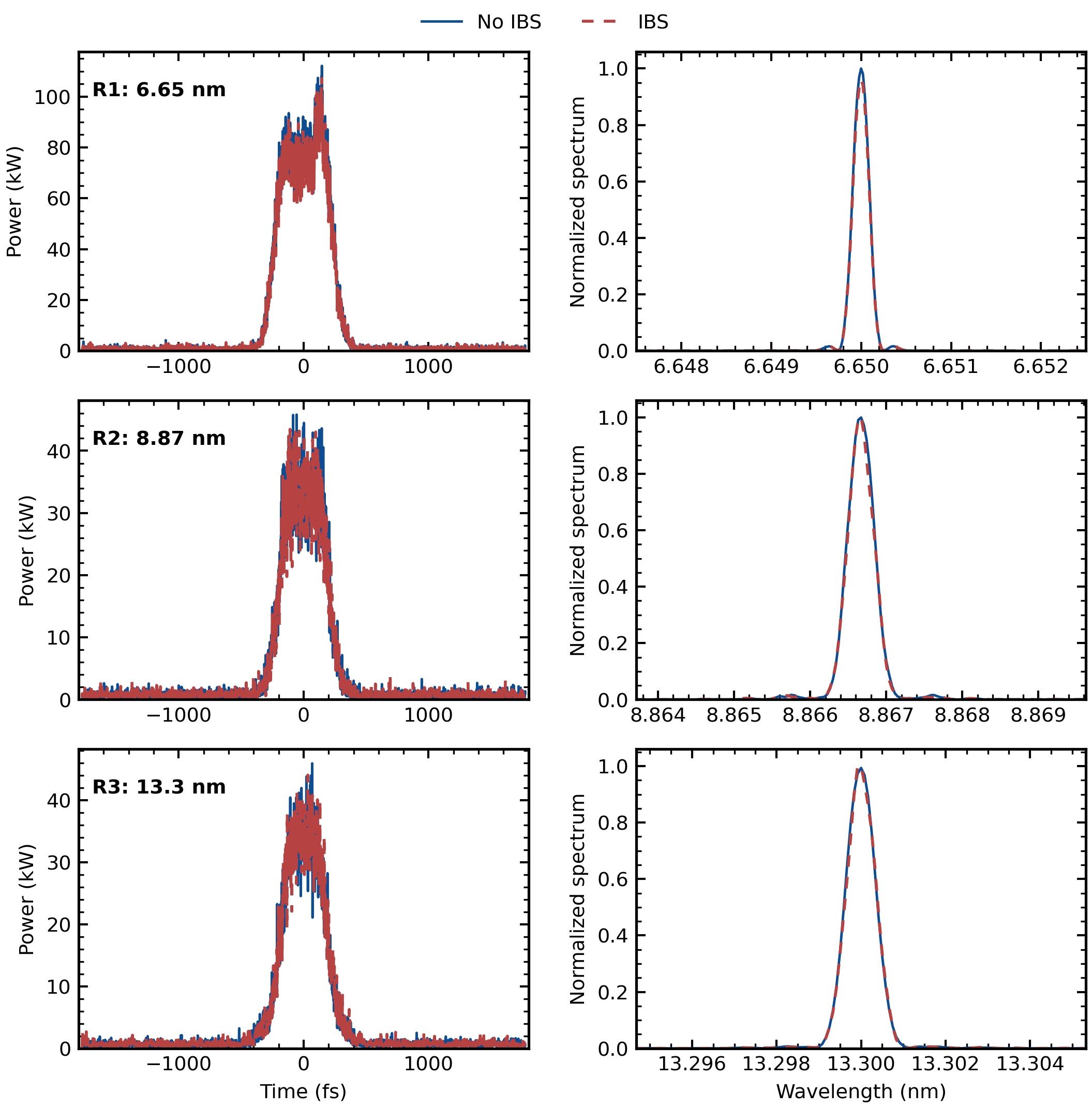}
 \caption{Single-pass IBS comparison.  Top: longitudinal energy bands after the first pair of arcs and before the first short chicane; the same longitudinal window and band-identification procedure are used in both cases.  The difference between the two panels therefore isolates the incremental IBS contribution, which broadens but does not merge adjacent bands.  Bottom: radiation power profiles and spectra from the complete downstream chains.  IBS slightly decreases the profile amplitudes while leaving the spectral locations and widths essentially unchanged at the present resolution.}
\label{fig:ibs_effects}
\end{figure*}

The downstream radiation is likewise only weakly affected.  Relative to the no-IBS case, the peak powers at 6.65, 8.87, and 13.3~nm are reduced by $4.42\%$, $5.17\%$, and $4.11\%$, respectively.  No resolvable shift in the central wavelength or change in the spectral full width at half maximum is observed for any of the three radiators.  Thus, the cumulative single-pass IBS contribution produces only a few-percent decrease in output power without disrupting coherent radiation generation at any stage.

\subsection{Self-consistent and cumulative coherent-radiation back-reaction}

Unlike the analytical map in Sec.~II, the numerical model retains the electron--radiation interaction in every radiator.  GENESIS advances the radiation field and electron equations of motion self-consistently \cite{Reiche1999GENESIS}, including energy exchange, ponderomotive-phase rotation, resonant-scale modulation, and slice-energy-spread growth.  The complete macroparticle distribution at each radiator exit is transferred to the following transport and echo stage.  The beam entering the second stage therefore contains the reaction from $R_1$, while that entering the third contains the cumulative reactions from $R_1$ and $R_2$; the $R_3$ reaction affects only the final radiator evolution and the spent beam.  The preservation of downstream bunching and coherent output in this sequential calculation demonstrates directly that the included cumulative back-reaction does not interrupt either subsequent echo.

The scale of this result can be understood with the spontaneous-self-interaction theory of a prebunched electron beam developed by Gover and co-workers \cite{Gover2019Superradiance}.  In a one-dimensional, single-mode description, the radiation amplitude $|\widetilde E|$, mean electron energy $\gamma$, and relative ponderomotive phase $\psi$ obey coupled equations of the form
\begin{equation}
 \frac{d|\widetilde E|}{ds}={\cal B}\sin\psi,\qquad
 \frac{d\gamma}{ds}=-{\cal C}|\widetilde E|\sin\psi,\qquad
 \frac{d\psi}{ds}\simeq-2k_u\frac{\gamma-\gamma_r}{\gamma_r}
 +\frac{{\cal B}}{|\widetilde E|}\cos\psi,
 \label{eq:superradiant_self_interaction}
\end{equation}
where ${\cal B}$ and ${\cal C}$ contain the beam--mode coupling and $\gamma_r$ is the resonant energy.  Equation~(\ref{eq:superradiant_self_interaction}) retains both energy exchange and ponderomotive-phase evolution and conserves the combined beam--radiation energy.  With no injected field, the field generated by the pre-existing bunching is initially in quadrature with it, so that $\psi\rightarrow\pi/2$ as $s\rightarrow0^+$.  Consequently, $|\widetilde E|\propto s$ and $P_\gamma\propto s^2$ at the radiator entrance, while the induced energy change and phase rotation enter successively as $\Delta\gamma\propto s^2$ and $\Delta\psi\propto s^3$.  A prebunched beam can therefore produce appreciable coherent radiation before significant self-induced phase-space rotation develops.

Energy conservation provides the corresponding numerical scale:
\begin{equation}
 U_{\gamma,i}=\frac{1}{e}\int dt\,I(t)\Delta E_{\gamma,i}(t),
 \qquad
 \left\langle\Delta E_{\gamma,i}\right\rangle_Q=\frac{eU_{\gamma,i}}{Q}.
 \label{eq:radiation_energy_balance}
\end{equation}
Coherent emission is concentrated within the approximately 400~fs, 32~A core, for which $Q_{\mathrm{eff}}\simeq I\Delta t=12.8$~pC.  The simulated pulse energies $42.68$, $16.90$, and $16.52$~nJ therefore correspond to core-averaged extracted-energy scales of $3.33$, $1.32$, and $1.29$~keV for $R_1$--$R_3$, respectively.  These values characterize the radiation-producing core and are not bounds on the microscopic energy modulation.

For a radiator, Eq.~(\ref{eq:superradiant_self_interaction}) gives the characteristic phase scale
\begin{equation}
 |\Delta\psi_{\gamma,i}|
 \lesssim 2k_{u,i}L_{u,i}\frac{|\Delta E_{\gamma,i}|}{E_0}
 =4\pi N_{u,i}\frac{|\Delta E_{\gamma,i}|}{E_0}.
 \label{eq:superradiant_phase_bound}
\end{equation}
Equation~(\ref{eq:superradiant_phase_bound}) deliberately applies the final core-averaged extraction over the full radiator length rather than allowing it to build from zero.  With $E_0=3.5$~GeV, $L_u=3.6$~m, and $N_u=90$, 80, and 72, it gives $1.08\times10^{-3}$, $3.79\times10^{-4}$, and $3.34\times10^{-4}$~rad for $R_1$--$R_3$.  If the initial coherent-emission scaling $\Delta E_\gamma(s)=\Delta E_{\gamma,\mathrm{exit}}(s/L_u)^2$ is used instead, then
\begin{equation}
 2k_u\int_0^{L_u}\frac{\Delta E_\gamma(s)}{E_0}\,ds
 =\frac{2k_uL_u}{3}\frac{\Delta E_{\gamma,\mathrm{exit}}}{E_0},
 \label{eq:superradiant_quadratic_phase}
\end{equation}
so the corresponding phase estimates are smaller by a factor of three.  The effect of these phase rotations can be related directly to the energy-exchange term in Eq.~(\ref{eq:superradiant_self_interaction}).  Because the coherently generated field is initially in quadrature with the bunching, a phase rotation $\Delta\psi$ changes the normalized energy-exchange factor from unity to $\cos(\Delta\psi)$.  The conservative phase bounds in Eq.~(\ref{eq:superradiant_phase_bound}) therefore correspond to fractional changes $1-\cos|\Delta\psi|$ of $5.8\times10^{-7}$, $7.2\times10^{-8}$, and $5.6\times10^{-8}$ for $R_1$--$R_3$, respectively.  Through the energy balance in Eq.~(\ref{eq:radiation_energy_balance}), these values also bound the phase-rotation-induced fractional changes in the extracted electron energy and, consequently, in the radiation pulse energy.  Accounting for the quadratic emission buildup in Eq.~(\ref{eq:superradiant_quadratic_phase}) reduces these changes by approximately a further factor of nine.  Thus, the phase-rotation correction to the coherent output is negligible on the scale relevant to the subsequent echoes.  Equations~(\ref{eq:superradiant_self_interaction})--(\ref{eq:superradiant_quadratic_phase}) serve only as one-dimensional scale estimates: the actual EEHG distribution has finite energy spread and a nontrivial bunching-phase distribution, while diffraction, slippage, resonant-scale structure, and cumulative transport are retained in the sequential \textsc{ELEGANT}--GENESIS simulations.  The analytic estimates and sequential simulations thus consistently show that coherent-radiation back-reaction does not disrupt the subsequent echoes.
}

\begin{acknowledgments}
This work was supported by National Natural Science Foundation of China (Nos. 12405176, and 12275284), Guangdong Basic and Applied Basic Research Foundation (No. 2026A1515030037), and Talent Program of Guangdong Province (No. 2024QN11X220).
\end{acknowledgments}


\bibliography{refs}

\end{document}